# Thermonuclear Reflect AB-Reactor*

**Alexander Bolonkin**
C&R, 1310 Avenue R, #F-6, Brooklyn, NY 11229, USA
T/F 718-339-4563, aBolonkin@juno.com, aBolonkin@gmail.com, http://Bolonkin.narod.ru

## Abstract

The author offers a new kind of thermonuclear reflect reactor. The remarkable feature of this new reactor is a three net AB reflector, which confines the high temperature plasma. The plasma loses part of its' energy when it contacts with the net but this loss can be compensated by an additional permanent plasma heating. When the plasma is rarefied (has a small density), the heat flow to the AB reflector is not large and the temperature in the triple reflector net is lower than 2000 - 3000 K. This offered AB-reactor has significantly less power then the currently contemplated power reactors with magnetic or inertial confinement (hundreds-thousands of kW, not millions of kW). But it is enough for many vehicles and ships and particularly valuable for tunnelers, subs and space apparatus, where air to burn chemical fuel is at a premium or simply not available.

The author has made a number of innovations in this reactor, researched its theory, developed methods of computation, made a sample computation of typical project. The main point of preference for the offered reactor is its' likely cheapness as a power source.

----------------------

*Presented to http://arxiv.org

## Introduction
### Brief information about thermonuclear reactors

**Fusion power** is useful energy generated by nuclear fusion reactions. In this kind of reaction two light atomic nuclei fuse together to form a heavier nucleus and release energy. The largest current experiment, JET, has resulted in fusion power production somewhat larger than the power put into the plasma, maintained for a few seconds. In June 2005, the construction of the experimental reactor ITER, designed to produce several times more fusion power than the power into the plasma over many minutes, was announced. The production of net electrical power from fusion is planned for the next generation experiment after ITER.

Unfortunately, this task is not easy, as scientists thought earlier. Fusion reactions require a very large amount of energy to initiate in order to overcome the so-called *Coulomb barrier* or *fusion barrier energy*. The key to practical fusion power is to select a fuel that requires the minimum amount of energy to start, that is, the lowest barrier energy. The best fuel from this standpoint is a one-to-one mix of deuterium and tritium; both are heavy isotopes of hydrogen. The D-T (Deuterium & Tritium) mix has a low barrier energy. In order to create the required conditions, the fuel must be heated to tens of millions of degrees, and/or compressed to immense pressures.

At present, D-T is used by two main methods of fusion: inertial confinement fusion (ICF) and magnetic confinement fusion (MCF)(for example, tokamak).

In **inertial confinement fusion** (**ICF**), nuclear fusion reactions are initiated by heating and compressing a target. The target is a pellet that most often contains deuterium and tritium (often only micro or milligrams). Intense laser or ion beams are used for compression. The beams explosively detonate the outer layers of the target. That accelerats the underlying target layers inward, sending a shockwave into the center of pellet mass. If the shockwave is powerful enough and if high enough density at the center is achieved some of the fuel will be heated enough to cause fusion reactions. In a target which has been heated and compressed to the point of thermonuclear ignition, energy can then heat surrounding fuel to cause it to fuse as well, potentially releasing tremendous amounts of energy.

Fusion reactions require a very large amount of energy to initiate in order to overcome the so-called *Coulomb barrier* or *fusion barrier energy*.

**Magnetic confinement fusion (MCF).** Since plasmas are very good electrical conductors, magnetic fields can also confine fusion fuel. A variety of magnetic configurations can be used, the basic distinction being between magnetic mirror confinement and toroidal confinement, especially tokamaks and stellarators.

**Lawson criterion**. In nuclear fusion research, the Lawson criterion, first derived by John D. Lawson in 1957, is an important general measure of a system that defines the conditions needed for a fusion reactor to reach **ignition**, that is, that the heating of the plasma by the products of the fusion reactions is sufficient to maintain the temperature of the plasma against all losses without external power input. As originally formulated the Lawson criterion gives a minimum required value for the product of the plasma (electron) density $n_e$ and the "energy confinement time" $\tau$. Later analyses suggested that a more useful figure of merit is the "triple product" of density, confinement time, and plasma temperature $T$. The triple product also has a minimum required value, and the name "Lawson criterion" often refers to this inequality.

The key to practical fusion power is to select a fuel that requires the minimum amount of energy to start, that is, the lowest barrier energy. The best fuel from this standpoint is a one-to-one mix of deuterium and tritium; both are heavy isotopes of hydrogen. The D-T (Deuterium & Tritium) mix has a low barrier.

**Short history of thermonuclear fusion**. One of the earliest (in the late 1970's and early 1980's) serious attempts at an ICF design was **Shiva**, a 20-armed neodymium laser system built at the Lawrence Livermore National Laboratory (LLNL) that started operation in 1978. Shiva was a "proof of concept" design, followed by the **NOVA** design with 10 times the power. Funding for fusion research was severely constrained in the 80's, but NOVA nevertheless successfully gathered enough information for a next generation machine whose goal was ignition. Although net energy can be released even without ignition (the breakeven point), ignition is considered necessary for a *practical* power system.

The resulting design, now known as the National Ignition Facility, commenced being constructed at LLNL in 1997. Originally intended to start construction in the early 1990s, the NIF is now six years behind schedule and overbudget by over $1.4 billion. Nevertheless many of the problems appear to be due to the "big lab" mentality and shifting the focus from pure ICF research to the nuclear stewardship program, LLNLs traditional nuclear weapons-making role. NIF is now scheduled to "burn" in 2010, when the remaining lasers in the 192-beam array are finally installed.

Laser physicists in Europe have put forward plans to build a £500m facility, called HiPER, to study a new approach to laser fusion. A panel of scientists from seven European Union countries believes that a "fast ignition" laser facility could make a significant contribution to fusion research, as well as supporting experiments in other areas of physics. The facility would be designed to achieve high energy gains, providing the critical intermediate step between ignition and a demonstration reactor. It would consist of a long-pulse laser with an energy of 200 kJ to compress the fuel and a short-pulse laser with an energy of 70 kJ to heat it.

Confinement refers to all the conditions necessary to keep a plasma dense and hot long enough to undergo fusion:

- **Equilibrium:** There must be no net forces on any part of the plasma, otherwise it will rapidly disassemble. The exception, of course, is inertial confinement, where the relevant physics must occur faster than the disassembly time.
- **Stability:** The plasma must be so constructed that small deviations are restored to the initial state, otherwise some unavoidable disturbance will occur and grow exponentially until the plasma is destroyed.
- **Transport:** The loss of particles and heat in all channels must be sufficiently slow. The word "confinement" is often used in the restricted sense of "energy confinement".

To produce self-sustaining fusion, the energy released by the reaction (or at least a fraction of it) must be used to heat new reactant nuclei and keep them hot long enough that they also undergo fusion reactions. Retaining the heat generated is called energy **confinement** and may be accomplished in a number of ways.

The hydrogen bomb weapon has no confinement at all. The fuel is simply allowed to fly apart, but it takes a certain length of time to do this, and during this time fusion can occur. This approach is called **inertial confinement** (fig.1). If more than about a milligram of fuel is used, the explosion would destroy the machine, so controlled thermonuclear fusion using inertial confinement causes tiny pellets of fuel to explode several times a second. To induce the explosion, the pellet must be compressed to about 30 times solid density with energetic beams. If the beams are focused directly on the pellet, it is called **direct drive**, which can in principle be very efficient, but in practice it is difficult to obtain the needed uniformity. An alternative approach is **indirect drive**, in which the beams heat a shell, and the shell radiates x-rays, which then implode the pellet. The beams are commonly laser beams, but heavy and light ion beams and electron beams have all been investigated and tried to one degree or another.

They rely on fuel pellets with a "perfect" shape in order to generate a symmetrical inward shock wave to produce the high-density plasma, and in practice these have proven difficult to produce. A recent development in the field of laser-induced ICF is the use of ultra-short pulse multi-petawatt lasers to heat the plasma of an imploding pellet at exactly the moment of greatest density after it is imploded conventionally using terawatt scale lasers. This research will be carried out on the (currently being built) OMEGA EP petawatt and OMEGA lasers at the University of Rochester and at the GEKKO XII laser at the Institute for Laser Engineering in Osaka Japan which, if fruitful, may have the effect of greatly reducing the cost of a laser fusion-based power source.

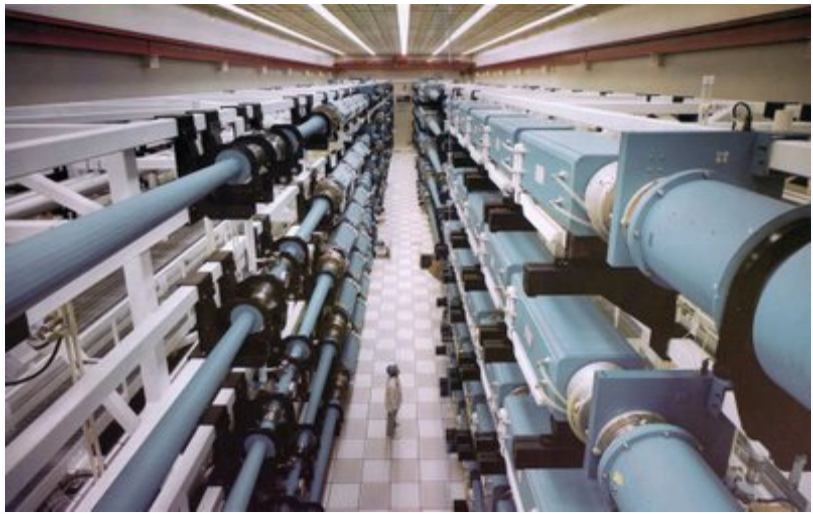

**Fig.1**. Laser installation for NOVA inertial thermonuclear reactor. Note the size of the man in comparison to the gigantic size of the class of laser installation needed for a reactor. Cost is some billions of dollars.

At the temperatures required for fusion, the fuel is in the form of a plasma with very good electrical conductivity. This opens the possibility to confine the fuel and the energy with magnetic fields, an idea known as **magnetic confinement** (fig.2).

Much of this progress has been achieved with a particular emphasis on tokamaks (fig.2).

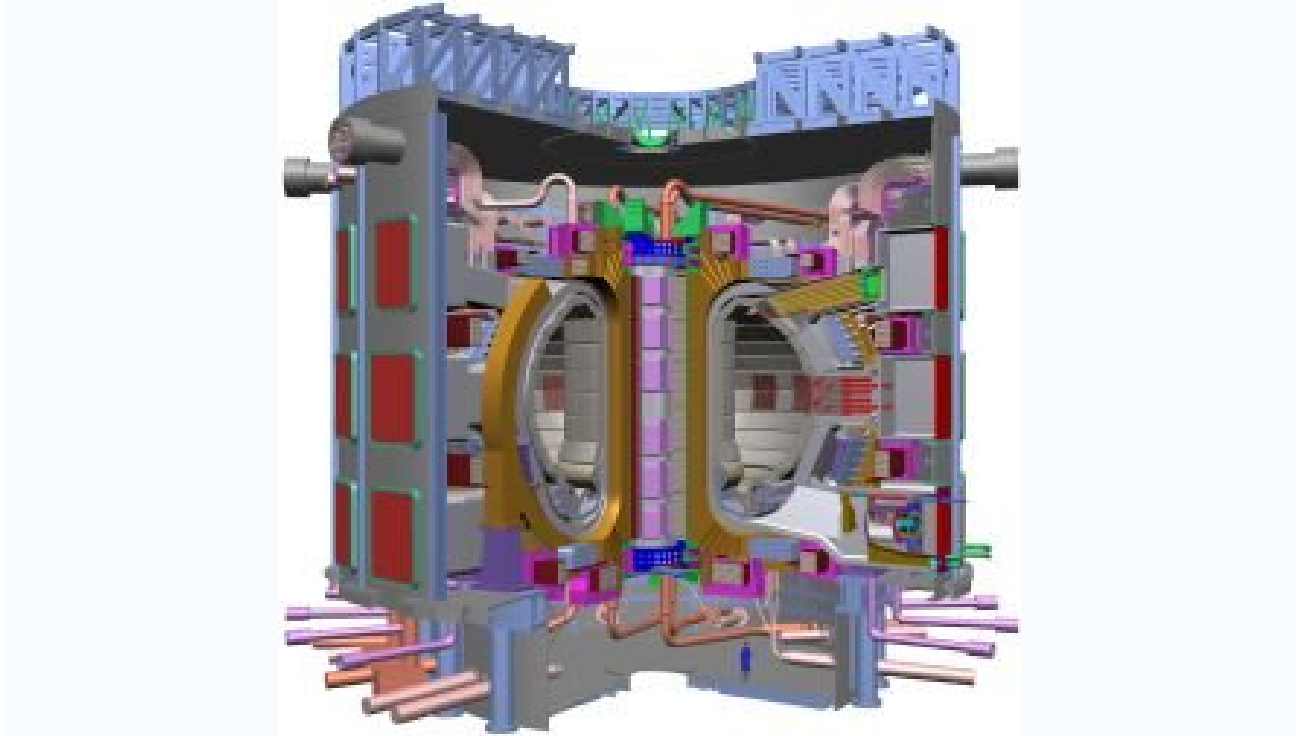

**Fig. 2**. Magnetic thermonuclear reactor (ITER). The size of the installation is obvious if you compare it with the "Little Blue Man" inside the machine at the bottom. Cost is $12.8-billions USD.

In fusion research, achieving a fusion energy gain factor $Q = 1$ is called **breakeven** and is considered a significant although somewhat artificial milestone. **Ignition** refers to an infinite $Q$, that is, a self-sustaining plasma where the losses are made up for by fusion power without any external input. In a practical fusion reactor, some external power will always be required for things like current drive, refueling, profile control, and burn control. A value on the order of $Q = 20$ will be required if the plant is to deliver much more energy than it uses internally.

In a fusion power plant, the nuclear island has a **plasma chamber** with an associated vacuum system, surrounded by a plasma-facing components (first wall and divertor) maintaining the vacuum boundary and absorbing the thermal radiation coming from the plasma, surrounded in turn by a blanket where the neutrons are absorbed to breed tritium and heat a working fluid that transfers the power to the balance of plant. If magnetic confinement is used, a **magnet** system, using primarily cryogenic superconducting magnets, is needed, and usually systems for heating and refueling the plasma and for driving current. In inertial confinement, a **driver** (laser or accelerator) and a focusing system are needed, as well as a means for forming and positioning the **pellets**.

The magnetic fusion energy (MFE) program seeks to establish the conditions to sustain a nuclear fusion reaction in a plasma that is contained by magnetic fields to allow the successful production of fusion power.

In thirty years, scientists have increased the Lawson criterion of the ICF and tokamak installations by tens of times. Unfortunately, all current and some new installations (ICF and totamak) have a Lawrence criterion that is tens of times lower than is necessary (fig.3).

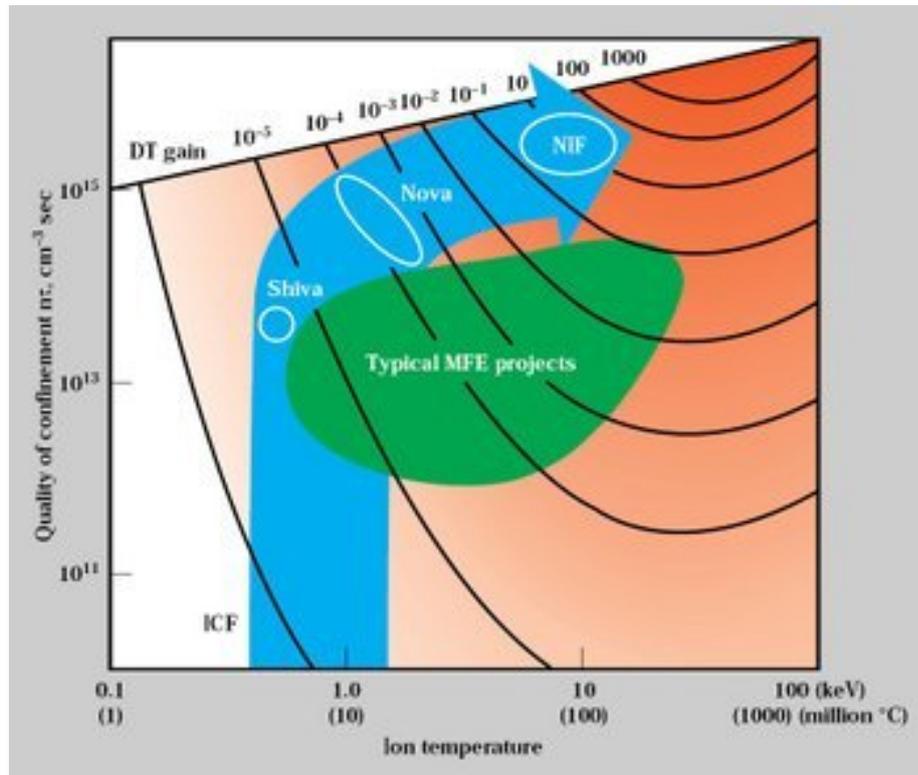

**Fig. 3.** Parameter space occupied by inertial fusion energy and magnetic fusion energy devices. Theregime allowing thermonuclear ignition with high gain lies near the upper right corner of the plot.

**Economics.** It is far from clear whether nuclear fusion will be economically competitive with other forms of power. The many estimates that have been made of the cost of fusion power cover a wide range, and indirect costs of and subsidies for fusion power and its alternatives make any cost comparison difficult. The low estimates for fusion appear to be competitive with but not drastically lower than other alternatives. The high estimates are several times higher than alternatives**.**

While fusion power is still in early stages of development, vast sums have been and continue to be invested in research. In the EU almost € 10 billion was spent on fusion research up to the end of the 1990s, and the new ITER reactor alone is budgeted at € 10 billion. It is estimated that up to the point of possible implementation of electricity generation by nuclear fusion, R&D will need further promotion totalling around € 60-80 billion over a period of 50 years or so (of which € 20-30 billion within the EU). In the current EU research programme (FP6), nuclear fusion research receives € 750 million (excluding ITER funding), compared with € 810 million for all non-nuclear energy research combined, putting research into fusion power well ahead of that of any single rivaling technology.

Unfortunately, despite optimism dating back to the 1950s about the wide-scale harnessing of fusion power, there are still significant barriers standing between current scientific understanding and technological capabilities and the practical realization of fusion as an energy source. Research, while making steady progress, has also continually thrown up new difficulties. Therefore it remains unclear that an economically viable fusion plant is even possible." An editorial in New Scientist magazine explained that "if commercial fusion is viable, it may well be a century away."

An important aspect of fusion energy in contrast to many other energy sources is that the cost of production is inelastic. The cost of wind energy, for example, goes up as the optimal locations are developed first, while further generators must be sited in less ideal conditions. With fusion energy, the production cost will not increase much, even if large numbers of plants are built. It has been suggested that even 100 times the current energy consumption of the world is possible.

Some problems which are expected to be an issue in the next century such as fresh water shortages can actually be regarded merely as problems of energy supply. For example, in desalination plants, seawater can be converted into pure freshwater through a process of either distillation or reverse osmosis. However, these processes are energy intensive. Even if the first fusion plants are not competitive with alternative sources, fusion could still become competitive if large scale desalination requires more power than the alternatives are able to provide.

Despite being technically non-renewable, fusion power has many of the benefits of long-term renewable energy sources (such as being a sustainable energy supply compared to presently-utilized sources and emitting no greenhouse gases) as well as some of the benefits of such much more finite energy sources as hydrocarbons and nuclear fission (without reprocessing). Like these currently dominant energy sources, fusion could provide very high power-generation density and uninterrupted power delivery (due to the fact that they are not dependent on the weather, unlike wind and solar power).

Several fusion reactors have been built, but as yet none has produced more thermal energy than electrical energy consumed. Despite research having started in the 1950s, no commercial fusion reactor is expected before 2050. The ITER project is currently leading the effort to commercialize fusion power.

**Summary.** At present time the most efforts of scientists are directed toward very large, superpower thermonuclear stations (Shiva, NOVA, NIF, LLNL, HiPER, OMEGA EP, ITER, Z-machine, etc.). These stations request gigantic finances, years of development, complex technology. That is well for scientists seeking a stable career path over many years, (generations!) but not well for the technical progress of humanity. Governments spent billions of dollars for development of thermonuclear technology. However, we have not had achieved a stable long duration thermonuclear reaction after *50 years* of thermonuclear development. In the author's opinion, industrial thermonuclear electric stations may appear after 20 – 40 more years and their energy will be more expensive than a current technology conventional electric station.

The author offers to direct government attention toward development smaller cheaper thermonuclear installations, which don't require huge funds and decades for development. Smaller power units are actually more practical in an immediate way: The world economy depends on transport vehicles (cars, ships, aviation), which are the main users of oil fuel. They very much need small thermonuclear engines. (Large fixed conventional stations use coal, replacement of which does not necessarily relieve pressure on world oil prices.)

Humanity needs new sources of energy. The offered innovations and researches are in [1]-[43].

## Description and Innovation

**1. AB plasma reflector**. A three net plasma reflector was offered by the author in his early works (see, for example, [5-10]). The plasma reflector (fig.4) has three conductive nets (*1, 2, 3*) and electric voltage between them. *V*oltage between nets *1* and *2* nets is about 100 kV in our case, when plasma temperature is 50 keV. The electric intensity is taken so that *1, 2* reflect the positive charged particles (for example atoms of deuterium D and tritium T). The voltage between the nets *2* and *3* nets is about 200 kV in our case. The electric intensity is taken so the *2, 3* reflect the negative charged particles (electrons).

The reflector works in the following way. When plasma is inserted in the reflector confinement, the nets *1, 2* reflect the positive charged atoms back to the plasma if the electric voltage is more than plasma temperature in eV. The particle speed has a Maxwellian distribution. That way the voltage between nets must be a minimum of two times more than the average temperature of the plasma. In this case, most positive particles will be reflected back to the plasma. Simultaneously, the nets *1,2* accelerate the negative particles (electrons) and increase their energy (temperature) additional up to 100 keV. That way the voltage between nets 2, 3 must be more then *V* plus the electron temperature of plasma (about 200 keV > 100 + 50 keV in our case).

The plasma contacts with the nets and heats them. This heating significantly depends upon the density of the plasma. For example, the Earth's atmosphere over 160 km has temperature more than 1200 K. but Earth satellites fly some years at this atmosphere without any perceptible heating.

The inner Van Allen Belt extends from an altitude of 700–10,000 km (0.1 to 1.5 Earth radii) above the Earth's surface, and contains high concentrations of energetic protons with energies exceeding 100 MeV and electrons in the range of hundreds of kiloelectronvolts, trapped by the strong (relative to the outer belts) magnetic fields in the region.

The charged particles in Earth's radiation belts have a temperature of some billions degrees, but a space apparatus orbiting there has a low temperature. It is possible because the density of plasma at these altitudes is very low and a heat flow from the high temperature plasma is small. This effect is used in the offered thermonuclear AB-reactor.

We use a plasma which has a density of a million times less than atmospheric air and a hundred times less than a conventional tokomak. As the result, our net has an equilibrium temperature (after its' own re-radiation of heat) less than the melting temperature of a refractory conductive material (for example, tungsten has a melting temperature of 3416 C = 3689 K).

Low density plasma produces low energy (a low power output). The current thermonuclear reactors under development are contemplated to cost milliards (billions of) USA dollars and would be reduced to operational practice as power electric stations having some millions of kW capacity. Therefore, this class of reactor, with a lower plasma density, is not acceptable for them. The authors' opinion is that such powerful thermonuclear stations would appear on the market not earlier than after 30 – 40 years. (Merely demonstrating breakeven does not means we get practical thermonuclear energy. There are a lot of problems that we must solve before industrial-scale production of thermonuclear energy becomes practical: For example, protection of people from neutrons, converting the neutron energy into electricity, etc. I am sure the big thermonuclear energy for many years will be more costly then conventional heat energy. For example, the first conventional (uranium) nuclear energy was released about 50 years ago, but for years it was more expensive than unsubsidized coal power. But with this practical, buildable new approach we should be able to develop a usable thermonuclear reaction within 3 – 5 years.

The developing the small power thermonuclear installations (hundreds and thousands kW, as offered by author) is useful for vehicles, sea ships, and space apparatus and will help to solve many future problems of the big thermonuclear stations on a smaller, more realizable scale. Developing small reactors should cost thousands of times less than building ITER or NOVA.

The computations show the main contribution to net heating may not be contact of the plasma to the net, but a product of the thermonuclear reaction – neutrons (14 MeV) and alpha particles (3.5 MeV).
The neutrons have full reflection if the angle between trajectory and surface is 10 – 12 degrees. That way the author offers to use the cross-section (wedge) form of net wire (fig. 2b, form 2). This form will reflect the part of high energy particles and has more wire surface for thermal re-radiation.

We can also use the form fig.2b, 3 which has an internal channel for cooling liquid.

The second heating of net (less, than contact or thermonuclear particles) is produced by the Bremsstrahlung, or braking radiation. The contact heating is about 5 – 25% of the full energy.

Note that most heating occurs at the first net, which contacts with the nuclear particles. The second net only has contact with electrons. The mass of electrons and density of electron gas is less by about 2000 - 5000 times, than the mass of nuclear particles D, T. (The Deuterium and Tritium, or mix of heavy hydrogen isotopes that serve as nuclear fuel) That way we cannot be troubled by heating of the second net. The third net heats only by reaction product ($\alpha$, n) and Bremsstrahlung radiation.

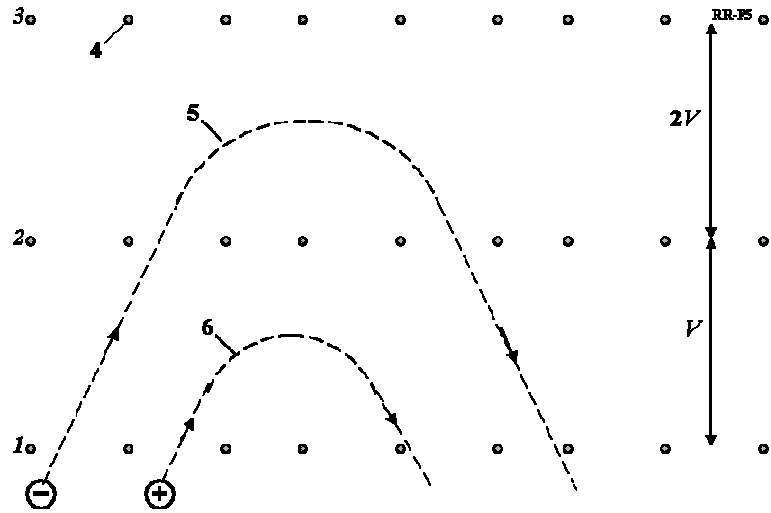

**Fig. 4.** Three net plasma AB-reflector. *Notation*: *1, 2, 3* – the first, second, and third nets; *4* - wire of net; *5* – trajectory of negative charged particles; *6* – trajectory of positive charged particles. *V* is voltage between nets *1* and *2* nets (about 100 kV in our case). The electric intensity is taken so the *1, 2* reflect the positive charged particles (for example, atoms D and T). *2V* is the voltage between the nets *2* and *3* nets (about 200 kV in our case). The electric intensity is taken so the *2, 3* reflect the negative charged particles (electrons).

We can also use hard material for nets, for example, tungsten, having a melting temperature of 3416 C. One tungsten atom has an atomic mass of about 184. The deuterium D has the atomic mass of 2 and tritium T has an atomic mass of 3. That means the light atoms hit on the heavy atom and passes to the heavy atom only a small part of its kinetic energy. The other important requirement of the net material must be a good tensile stress in high temperature. The tungsten has such a good property mix. Carbide of tantalum and zirconium has the melt temperature up to 3500 – 3900 C. The nanotube has good stability up to 2300 C in vacuum and excellent tensile stress.

Innovations in the plasma reflector in comparison with a conventional ' particle mirror' are the following:
1. The AB plasma reflector has three nets and it can reflect the plasma. Conventional electrostatic mirrors have two nets and it can reflect only charged particles of one sign.
2. Net wires can have an internal cooling (fig. 5b-3).
3. Cross-section of net wires can have a special form (fig. 5b-2) for reflecting the particles.
4. Net wires are made from heat-resistant electro-conductive material having good tensile stress in high temperature.
5. These net wires can have be electronically cooled.

**2. Space thermonuclear AB reactor**. Scheme of this reactor is presented in fig.5. This incarnation uses only charged α-particles and lets out (sets free) the neutrons and Bremstrahlung (X-soft ray) radiation (it doesn't protect from them). That significantly simplifies the reactor and decreases its mass. But this design can be used in space apparatus when the reactor is located separately (and far) from space ship (fig. 7). The manned space cabin can have additional light protection.

The reactor has the spherical form and contains: AB spherical reflector enclosed D+T plasma, spherical conductivity cover, two source of electric voltage (about 100 kV and 200 kV), outer electric user, plasma injector, and plasma electric heater.

Spherical conductivity cover is a thin (0.1 mm) aluminum film which collects the positive charged high energy (3.5 MeV) (α) particles, created by thermonuclear reaction. When they move between net 3 and reactor cover (3, fig. 5), they are braked and accumulated by the cover. As the result the thermonuclear reactor produces electric current-- about 3 MV in voltage. Part of this energy is used for heating the plasma, fuel, and support of the voltage in plasma reflector.

The loss of plasma temperature from contact plasma with reflector net and Bremstrahlung (X-ray) radiation is compensated by a plasma heater (fig. 5, notation 7) by passing the electric current through the plasma.

The reactor works continuously and produce tens to hundreds of kW of electric energy. That output harnesses only 20% of the liberated thermonuclear energy. (Efficiency is sacrificed for the sake of lightness, vital for space engines.)

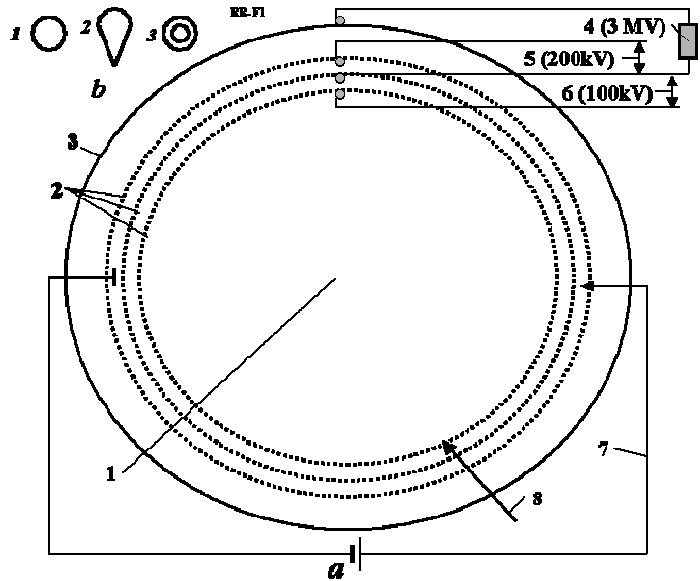

**Fig.5.** Scheme of Space Thermonuclear reflector AB-reactor, having spherical form.
*Notations*: a) Reactor: 1 – plasma; 2 – three net AB-reflector; 3 – conductivity cover; 4 – voltage between net and cover; 5 – voltage between the second and third nets; 6 – voltage between the first and the second net; 7 – plasma electric heater; 8 – fuel injector. b) Cross section of the net wire. *Notation*: 1 – round; 2 – spherical (wedge) form; 3 – tube.

**3. Earthbound thermonuclear reactor**. The space reactor is not acceptable for Earth's biosphere because that version produces neutrons and X-ray radiation. The neutrons create radioisotopes which damage biological creatures. The Earthbound reactor needs a special protection for men and environment from neutrons and Bremsstrahlung radiation. This protection increases the reactor mass by some (4-8) times. But that allows utilizing the full energy of reactor and produces the tritium – the second important component of thermonuclear fuel. The full energy received from the thermonuclear reactor increases by 4 times. About 40% this additional energy can be converted to electric energy (total is 60%) and the rest may be utilized for home heating, production of freshwater, etc.

The offered Earthbound version of the thermonuclear reactor is shown in fig. 6. Reactor contains: installation of fig. 5 inserted into protection cover. Protective cover has cooling tubes 9, which connects to vapor turbine 11. The turbine drives the electric generator 12, producing useful energy. The vapor after turbine flows to the heat exchanger 13 which produces hot water (vapor) 14 for home heating (freshwater production, etc.).

The reactor has lithium blankets 10 for reproduction of tritium.

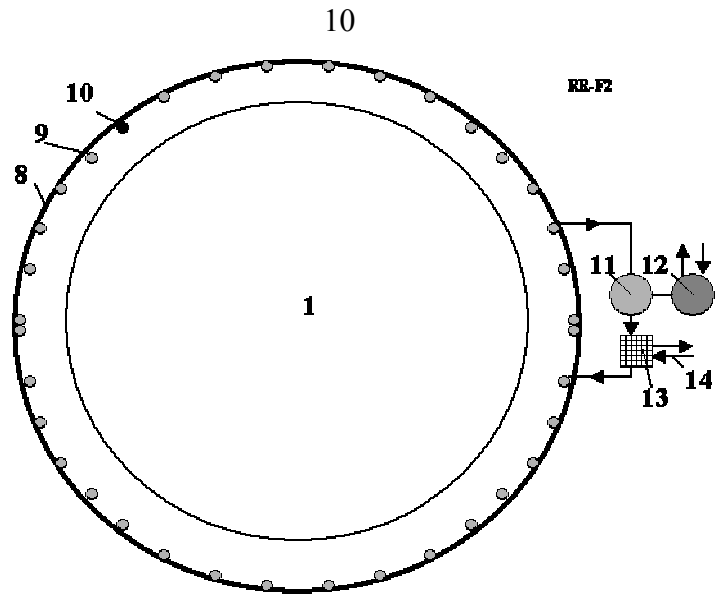

**Fig. 6**. Scheme of Earth's thermonuclear reflector AB-reactor. *Notations*: 1 – reactor of fig. 5; 8 – protection against neutrons (neutron moderator); 9 – cooling (turbine) tubes; 10 – blanket for production of tritium; 11 - vapor turbine; 12 – electric generator; 13 – heat exchanger; 14 – hot water (vapor) for home heating (freshwater production, etc.).

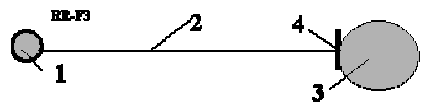

**Fig. 7.** Reflector AB-reactor in space. *Notation*: 1 – space reactor; 2 – connection to space ship; 3 – space ship; 4 – protection from neutrons and reactor radiation.

**4. Thermonuclear fuel.** The reactor fuel is deuterium and tritium, D + T. This reaction needs a high temperature about 1 -5 keV.

**Deuterium** (symbol D), also called **heavy hydrogen**, is a stable isotope of hydrogen with a natural abundance in the oceans of Earth of approximately one atom in 6500 of hydrogen. Deuterium thus accounts for approximately 0.015% (on a weight basis, 0.030%) of all naturally occurring hydrogen in the oceans on Earth. The nucleus of deuterium, called a **deuteron**, contains one proton and one neutron. Deuterium occurs in trace amounts naturally as deuterium gas, written $^2H_2$ or $D_2$, Density: 0.180 kg/m³ at STP (0 °C, 101.325 kPa). Data at approximately 18 K for $D_2$ (triple point): Density: liquid - 162.4 kg/m³, gas - 0.452 kg/m³. Viscosity: 12.6 µPa·s at 300 Kelvin (gas phase). Specific heat capacity at constant pressure $c_p$: solid 2950 J/(kg·K), gas 5200 J/(kg·K)

**Tritium** (symbol **T** or $^3H$) is a radioactive isotope of hydrogen. It is a gas ($T_2$ or $^3H_2$) at standard temperature and pressure. Tritium combines with oxygen to form a liquid called tritiated water $T_2O$ or partially tritiated THO. Tritium occurs naturally due to cosmic rays interacting with atmospheric gases. Because of tritium's relatively short half-life, however, tritium produced in this manner does not accumulate over geological timescales, and its natural abundance is negligible. Industrially, tritium is produced in nuclear reactors by neutron activation of lithium-6. Tritium is also produced in heavy water-moderated reactors when deuterium captures a neutron; however, this reaction has a much smaller cross section and is only a useful tritium source for a reactor with a very high neutron flux. It can also be produced from boron-10 through neutron capture. The tritium has half-life time only 12.32 years. That way the tritium is absent in the Earth. One is in Moon in small amount because the Moon doesn't have atmosphere and cosmic rays reach its surface and produce the tritium.

Current nuclear fusion research is focused on the D + T thermonuclear fusion reaction
$$D + T \rightarrow {}^4He(3.5 \text{ MeV}) + n \text{ (14.1 MeV)},$$

This reaction can occur in high-temperature deuterium-tritium plasma. Most energy released by the reaction is converted to the kinetic energy of the neutron. Since the neutron is not confined or reflected by a magnetic or electrostatic field it leaves, going outwards to surrounding space or hits the screen or vessel wall (or blanker) immediately after reaction. In last instance, the neutron kinetic energy is converted to heat. The heat is taken away from the screen by direct radiation or and indirect circulating coolant and can be used to run an electric generator. If we add $^6$Li inside the blanket, then tritium can be produced by reaction

$$n + {}^6Li \rightarrow {}^4He\ (2.1\ MeV) + T\ (2.7\ MeV)$$

and then used as the fuel. Another reaction product is the alpha (α) particles $^4$He carrying 3.5 MeV which can be directed or confined by electro-magnetic field.

The reaction that produces only charged particles are best for the proposed propulsion system and generator. Unfortunately, these reactions are not great (see Table 1). All of them request the very high temperature for ignition and has the low cross-section of reaction (they produced low energy in volume unit).

**Table 1**. Aneutronic reactions.

| |
|---|
| D+$^3$He→$^4$He(3.6)+p(14.7) |
| D+$^6$Li→2$^4$He(22.4) |
| $^3$He+$^3$He→$^4$He(4.3)+2p(8.6) |
| $^3$He+$^6$Li→2$^4$He(1.9)+p(16.9) |
| p+$^{11}$B→3$^4$He(8.7) |
| p+$^6$Li→$^4$He(1.7)+$^3$He(2.3) |
| p+$^7$Li→2$^4$He(17.2) |

The **D-T** reaction is favored since it has the largest fusion cross-section (~ 5 barns peak) and reaches this maximum cross-section at the lowest energy (~ 65 keV center-of-mass) of any potential fusion fuel.

According to IEER's 1996 report about the United States Department of Energy, only 225 kg of tritium has been produced in the US since 1955. Since it is continuously decaying into helium-3, the stockpile was approximately, 75 kg at the time of the report.

# Theory, Computations, and Estimations

The estimation and computation of the offered AB-reactor are made better in this way:
Assign the radius of the first net $r = 0.5 – 10$ m, plasma pressure $n = 10^{18} – 10^{20}$ 1/m$^3$, and temperature of plasma $T = 1 – 50$ keV. *Note*: the plasma density is about $n = 10^{21}$ 1/m$^3$ in a conventional big tokomak.

For our example, let us take: $r = 3$ m, $n = 10^{19}$ 1/m$^3$, $T = 50$ keV $= 5 \times 10^4 \times 1.16 \times 10^4 = 5.8 \times 10^8$ K.

1. The surface and volume of a sphere are:
$$S = 4\pi r^2,\quad V = \frac{4}{3}\pi r^3,\quad S = 113\ \text{m}^2,\quad V = 113\ \text{m}^3, \tag{1}$$

2. Thermonuclear energy of reaction D+T for $T = 50$ keV released in form of charged particles is (in 1 m$^3$):
$$P_{DT} = 5.6 \cdot 10^{-13} 0.25 \cdot n^2 (\overline{\sigma v})_{DT} = 1.4 \cdot 10^{-13} \cdot 10^{26} 8.7 \cdot 10^{-16} = 1.22 \cdot 10^4\ \text{W/m}^3, \tag{2}$$

where $n$ is 1/cm$^3$ and $(\overline{\sigma v})_{DT}$ is taken from Table 2 below:

**Table 2.** Reaction rates $(\overline{\sigma v})_{DT}$ (in cm$^3$s$^{-1}$), averaged over Maxwellian distribution

| Tempera-ture, keV | Reaction D - D | Reaction D - T | Reaction D – He$^3$ |
|---|---|---|---|
| 5 | 1.8×10$^{-19}$ | 1.3×10$^{-17}$ | 6.7×10$^{-21}$ |
| 10 | 1.2×10$^{-18}$ | 1.1×10$^{-16}$ | 2.3×10$^{-19}$ |
| 20 | 5.2×10$^{-18}$ | 4.2×10$^{-16}$ | 3.8×10$^{-18}$ |
| 50 | 2.1×10$^{-17}$ | 8.7×10$^{-16}$ | 5.4×10$^{-17}$ |
| 100 | 4.5×10$^{-17}$ | 8.5×10$^{-16}$ | 1.6×10$^{-16}$ |

Source: AIP, 3-rd Edition, p.644.

The additional reaction noted in Table 2 are:

$$D + D \rightarrow {}^3H + {}^1H + 4.033 \text{ MeV } 50\%,$$
$$\rightarrow {}^3He + n + 3.27 \text{ MeV } 50\%, \quad (3)$$
$$D + He^3 \rightarrow {}^4He + {}^1H + 18.354 \text{ MeV},$$
$$\rightarrow {}^6Li + \gamma + 16.388 \text{ MeV},$$

The power density released in the form charged particle is:

$$P_{DD} = 3.3 \cdot 10^{-13} n_D^2 (\overline{\sigma v})_{DD}, \quad P_{DHe^3} = 2.9 \cdot 10^{-12} 0.25 \cdot n^2 (\overline{\sigma v})_{DHe^3}, \text{ W/cm}^3, \quad (4)$$

where $n$ is 1/cm$^3$ and $(\overline{\sigma v})_{DT}$ is taken from Table 2.

Full thermonuclear energy is

$$P_{DT,F} = P_{DT} V = 1.22 \cdot 10^4 \cdot 113 = 1.38 \cdot 10^6 \text{ W}. \quad (5)$$

3. Energy from surface 1 m$^2$:

$$P_{DT,S} = r P_{DT}/3 = 1.22 \cdot 10^4 \text{ W/m}^2. \quad (6)$$

4. Number reaction in 1 m$^3$

$$N = \frac{P_{DT}}{17.5 \text{ MeV}} = \frac{1.22 \cdot 10^4}{17.5 \cdot 10^6 \cdot 1.6 \cdot 10^{-19}} = 4.36 \cdot 10^{15} \text{ 1/m}^3, \quad (7)$$

5. Fuel consumption (D+T) by volume 1 m$^3$ in day

$$C_{f,1} = N \cdot m_i = 4.35 \cdot 10^{15} \cdot 5 \cdot 1.67 \cdot 10^{-27} =$$
$$= 1.82 \cdot 10^{-7} \text{ kg/s} \cdot \text{m}^3 = 0.157 \cdot 10^{-2} \text{ g/day} \cdot \text{m}^3, \quad (8)$$

Full fuel consumption is $C_f = C_{f,1} \times V$.

6. Expense power for fuel heating

$$p_f = NVkT = 4.36 \cdot 10^{15} \cdot 113 \cdot 1.38 \cdot 10^{-23} \cdot 5.8 \cdot 10^8 = 3.81 \cdot 10^3 \text{ W}, \quad (9)$$

where $k = 1.38 \times 10^{-23}$ is Bolzmann's constant, J/K.

7. Computation of reflector net.

a) Plasma pressure

$$p = nkT = 10^{19} \cdot 1.38 \cdot 10^{-23} 5.8 \cdot 10^8 = 8 \cdot 10^4 \text{ N/m}^2. \quad (10)$$

b) Pressure in one net is

$$p_1 = \frac{1}{3} p = 2.67 \cdot 10^4 \frac{N}{m^2}. \quad (11)$$

c) Thickness of net as continues cover

$$\delta = \frac{r p_1}{2\sigma} = \frac{3 \cdot 2.67 \cdot 10^4}{2 \cdot 4 \cdot 10^8} = 10^{-4} \text{ m} = 0.1 \text{ mm}, \quad (12)$$

where $\sigma = 4 \times 10^8$ N/m$^2$ = 40 kg/mm$^2$ is safety tensile stress of net wire.

d) Diameter net wire for net cell $L = 100 \times 100$ mm (Distance between nets is 0.5 m) is

$$D = 2\sqrt{\frac{\delta L}{\pi}} = \sqrt{\frac{0.1 \cdot 100}{3.14}} = 3.57 \text{ mm}. \quad (13)$$

e) Transparency of net is

$$\xi = \frac{2D}{L} = \frac{2 \cdot 3.57}{100} = 0.0714. \quad (14)$$

Coefficient of the net transparency is a very important value for computation of the reactor efficiency. If net transparency is less, the loss of plasma energy is smaller. If net transparency is high, the required compensation energy may be more than the received thermonuclear energy. We can decrease it by using a conductive material having a high safety tensile stress at high temperature. We can also increase the size $L$ of the net cell. But this method requires some increase in the size of reactor.

8. Volume Bremsstrahlung radiation of 1 m³ is
$$P_B = 5.34 \cdot 10^{-37} n^2 T^{0.5} = 5.34 \cdot 10^{-37} \cdot 10^{38} \cdot 50^{0.5} = 3.78 \cdot 10^2 \ \text{W/m}^3. \tag{15}$$
Here $n$ is in 1/m³, $T$ is in keV.
Total Bremsstrahlung radiation is
$$P_{B,T} = P_B V = 3.78 \cdot 10^3 \cdot 113 = 4.27 \cdot 10^4 \ \text{W}. \tag{16}$$
Surface Bremsstrahling radiation of 1 m² is
$$P_{B,S} = r \cdot P_B / 3 = 3.78 \cdot 10^2 \ \text{W/m}^2. \tag{17}$$

9. Contact heat transfer is
$$P_c = \frac{\rho}{\rho_a} k_1 T = 2 \times 10^3 \ \frac{\text{W}}{\text{m}^2}, \quad \text{where} \quad \rho = n\mu m_p, \quad \rho_a = n_a \mu_a m_p. \tag{18}$$
Here $\rho$ is density of plasma, kg/m³; $\rho_a$ = 1.225 kg/m³ is standard density of atmosphere air; $k_1$ = 100 W/m²K – heat transfer coefficient from gas to a solid well (one is right for $T$ equals some thousands K, but comparing with other estimations shown that may be applied for $T$ millions K); $\mu = m_i/m_p \approx 30$ for air and $\mu \approx (2 + 5)/2 = 2.5$ for D + T plasma. For our example, $P_c = 2 \times 10^3$ W/m².

10. Full energy fall into 1 m² of net surface is (sum of Bremsstrahlung radiation + plasma contact transfer)
$$q = \frac{P_{B,S}}{\pi} + 2 \cdot P_c = \frac{3.78 \cdot 10^2}{3.14} + 2 \cdot 2 \cdot 10^3 = 4 \cdot 10^3 \ \frac{\text{W}}{\text{m}^2}. \tag{19}$$

11. Temperature of the first net is
$$T = 100 \sqrt[4]{\frac{q}{C_S}} = 517 \ \text{K}, \quad \text{where} \quad C_S = 5.67 \ \frac{\text{J}}{\text{m}^2 \text{K}^4} - \text{heat coefficient.} \tag{20}$$

12. Energy is getting by Space reactor from electric generator for the efficiency coefficient $\eta$ = 0,9 :
$$E_e = \eta \cdot V \cdot P_{DT} = 0.9 \cdot 113 \cdot 1.22 \cdot 10^4 = 1,24 \cdot 10^6 \ \text{W}. \tag{21}$$

13. Less of energy for heating 3 nets is:
$$E_N = 3\xi \cdot V P_{DT} = 3 \cdot 0.0714 \cdot 113 \cdot 1.22 \cdot 10^4 = 2.95 \cdot 10^5 \ \text{W}. \tag{22}$$

14. Total less for supporting the continuous thermonuclear reaction (compensation of plasma loss for heating nets, fuel and radiation loss):
$$L_s = P_{B,T} + E_N + p_f = 4.27 \cdot 10^4 + 2.95 \cdot 10^5 + 0.381 \cdot 10^4 = 3.38 \cdot 10^5 \ \text{W}. \tag{23}$$

15. Useful electric power is:
$$P = E_e - L_s = 1.24 \cdot 10^6 - 3.38 \cdot 10^5 = 0.902 \cdot 10^6 \ \text{W} = 902 \ \text{kW} \tag{24}$$

16. Mass of Space reactor is:
   a) Mass of three nets
$$M_1 = 3\gamma \delta S = 3 \cdot 19.34 \cdot 10^3 \cdot 10^{-4} \cdot 113 = 656 \ \text{kg}. \tag{25}$$
Here $\gamma$ = 19340 kg/m³ is density of tungsten.
   b) Mass of aluminum cover having $\delta_2$ = 0.1 mm, $\gamma_2$ = 2700 kg/m³ is
$$M_2 = \gamma_2 \delta_2 S = 2.7 \cdot 10^3 \cdot 10^{-4} 113 \cdot 1.2 = 36.6 \ \text{kg}. \tag{26}$$
The radius of cover is about 4.2 m. Here coefficient 1.2 increases the mass.
   c) Total mass of Space reactor is about 700 – 800 kg.

17. Powerful power of Earth's version of AB Reactor is significantly more.
$$P = \frac{17.5}{3.5} P_{DT,F} - p_f = 5(1.38 \cdot 10^6 - 3.81 \cdot 10^3) \approx 6.9 \cdot 10^6 \ \text{W} = 6.9 \ \text{MW}. \tag{27}$$

In this version there is only the less for fuel heating. The neutron, Bremsstrahlung and nets radiation not leave the reactor. About 55 – 60% of this energy may be utilized as electric energy. The rest may be used as heat energy.

The mass of Earth's version is about 6 – 8 tons.

The results of computation via reactor radius for different plasma density are presented in figures 9 – 11. Fig. 8 shows the maximal tensile stress of the tungsten via temperature.

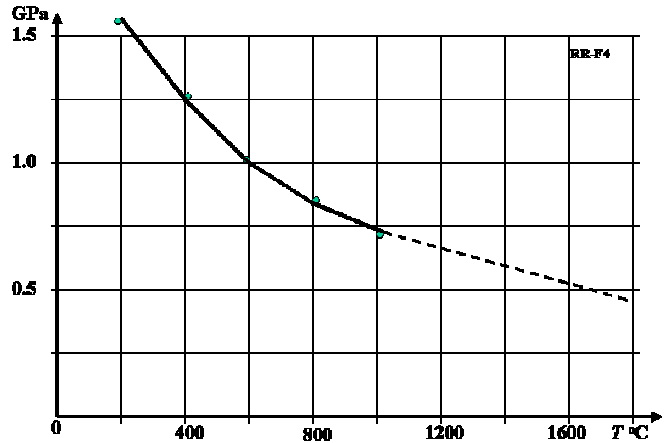

**Fig. 8**. Tensile stress of tungsten via temperature in C. Continuous curve is experiment, broken curve is extrapolation to the melting point $3416^\circ$ C.

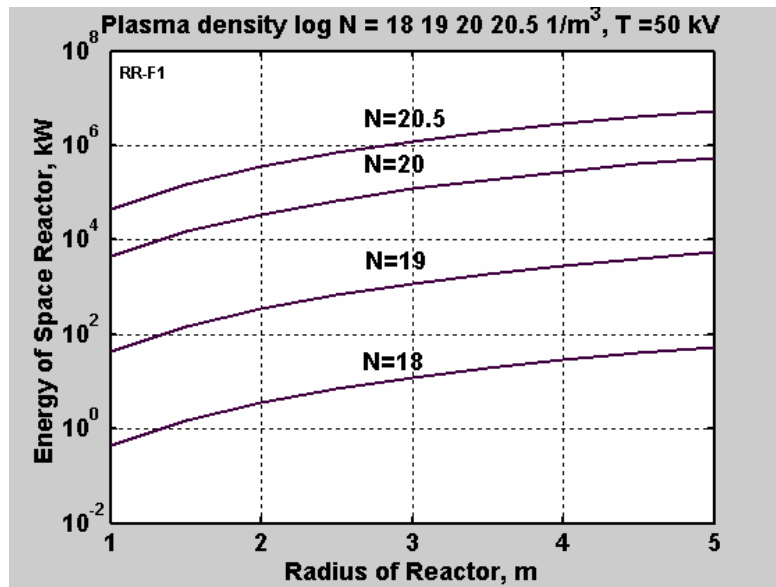

**Fig. 9**. Electric power of Thermonuclear reflector AB-reactor (18% from total power) via radius of reactor and plasma density. Initial data for computation: coefficient electric efficiency $\eta = 0.9$;

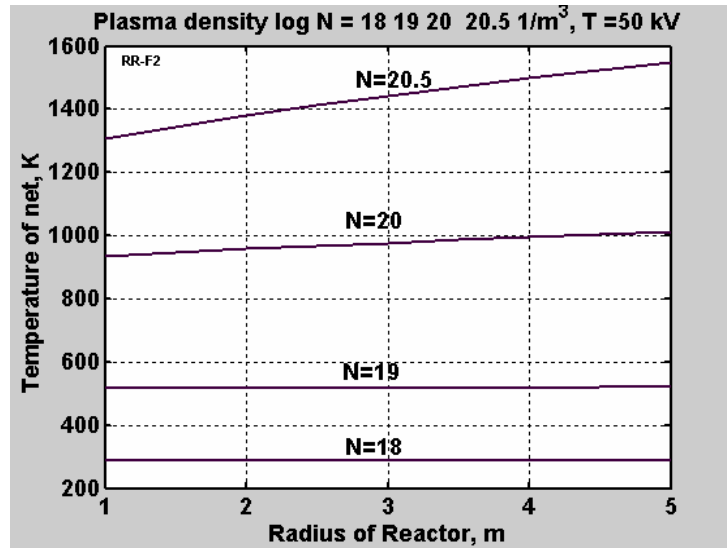

**Fig.10.** Temperature (K) of the first net via reactor radius and plasma density. Initial data for computation: plasma temperature 50 keV; net cell size 100 mm, form of net wire is round; specific density of tungsten is 19340 kg/m$^3$,
safety tensile stress of net for $T$ = 1500 K is $\sigma$ = 2×10$^8$ N/m$^2$.

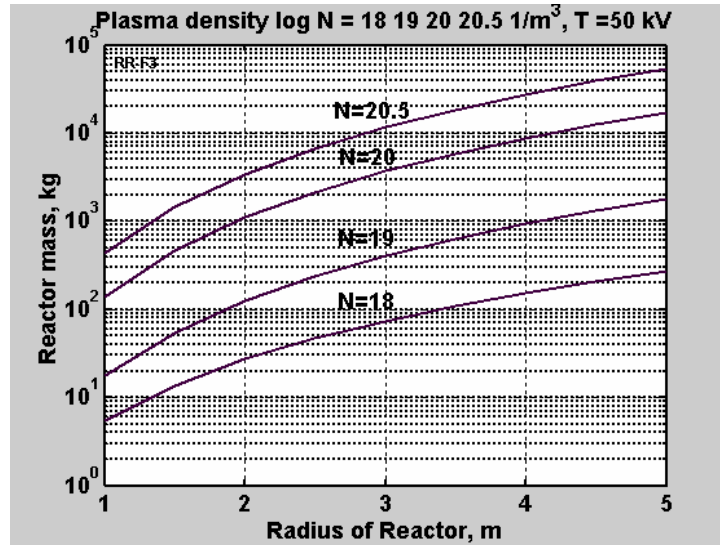

**Fig.11**. Reactor mass (kg) versus reactor radius and plasma density. Initial data for computation: plasma temperature 50 keV; net cell size $L$ = 100×100 mm, form wire is round; specific density of tungsten is $\gamma_1$ = 19340 kg/m$^3$, safety tensile stress of net for $T$ = 1500 K is $\sigma$ = 4×10$^8$ N/m$^2$; specific density of aluminum is $\gamma_2$ = 2700 kg/m$^3$, thickness of cover is 0.1 mm.

### Project.

As Space project of Thermonuclear AB Reflect Reactor may be used the computation above for initial data: $r$ = 3 m, $n$ = 10$^{19}$ 1/m$^3$, $T$ = 50 keV = 5.8×10$^8$ K. We get the useful electric power 902 kW and reactor mass about 400 kg. Full diameter of reactor is about 8.5 m. Distance between nets is about 0.5 m.

Selected reactor parameters are not best possible nor optimized. That is merely an example computation. The using of materials having more safety stress at high temperatures significantly decreases the net loss, (literally; the loss from the nets) and increases the useful electric power. The increasing of plasma density significantly increases the reactor power and decreases the reactor size (diameter) and mass.

The same notes apply to the Earthbound AB-reactor. That is more complex, has more mass, but it produces significantly more energy.

### Discussion

The low density plasma radically decreases the heat flow to solid surface. The high temperature solid surface intensive radiates the heat and balance may be lower than the melting temperature of solid conductive material. The entire magnetic bottle difficulty is avoided!

The confinement nets contact to a rare thermonuclear plasma and absorb a part of the plasma energy. This energy compensates (as X-radiation and heating of fuel) an electric energy produced by reactor. This total loss may be 5 – 30% of electric energy created by the charged $\alpha$ particles. If we compensate the loss of plasma energy by the electric (or high frequency) heating, we can receive a stable plasma which will permanently produce thermonuclear energy.

The disadvantage of the offered method is the high size of the reactor (from 1 through 8 m of diameter). That is consequence of a used low density plasma (low produced energy of rarefied plasma per volume) and required for a given engine power. But there are a lot of users who will be satisfied by the energy output of 100 – 2000 kW in a reactor 1 – 8 m of diameter.

At present time the most efforts of scientists are directed toward very large, superpower thermonuclear stations (Shiva, NOVA, NIF, LLNL, HiPER, OMEGA EP, ITER, Z-machine, etc.).

These stations request gigantic finances, years of development, complex technology. That is well for scientists seeking a stable career path over many years, (generations!) but not well for the technical progress of humanity. Governments spent billions of dollars for development of thermonuclear technology. However, we have not had achieved a stable long duration thermonuclear reaction after *50 years* of thermonuclear development. In the author's opinion, industrial thermonuclear electric stations may appear after 20 – 40 more years and their energy will be more expensive than a current technology conventional electric station.

The author offers to direct government attention toward development smaller cheaper thermonuclear installations, which don't require huge funds and decades for development. Smaller power units are actually more practical in an immediate way: The world economy depends on transport vehicles (cars, ships, aviation), which are the main users of oil fuel. They very much need small thermonuclear engines. (Large fixed conventional stations use coal, replacement of which does not necessarily relieve pressure on world oil prices.)

In the offered project, as in any innovation, the obstacles may appear (for example, in heating of the plasma). But the author has ideas for solution of these problems.

**Results**

Author offers a new (reflect reactor) confinement for thermonuclear low density plasma (not magnetic or inertial). He offered the series of innovations which allow applying this concept and develops the method for estimation and computation of a new class of small thermonuclear reactor. This reactor may be used for space, Earth transportation and small energy stations.

**Acknowledgement**

The author wishes to acknowledge Joseph Friedlander for correcting the author's English and for useful technical advice.